\documentclass[aps,twocolumn,superscriptaddress,showpacs,showkeys,floatfix]{revtex4}
\usepackage[bf]{caption}
\usepackage{graphicx}
\usepackage{amsmath}
\usepackage{wrapfig}

\begin{document}
 
\title{Information theory of open fragmenting systems}
\author{F.Gulminelli} %
\affiliation{LPC Caen (IN2P3 - CNRS / EnsiCaen et Universit\'{e}), F-14050
Caen C\'{e}dex, France} 
\altaffiliation{member of the Institut Universitaire de France}
\author{Ph.Chomaz}
\affiliation{GANIL ( DSM - CEA / IN2P3 - CNRS),B.P.5027, F-14076 Caen C\'edex 5, France}
\author{O.Juillet} %
\affiliation{LPC Caen (IN2P3 - CNRS / EnsiCaen et Universit\'{e}), F-14050
Caen C\'{e}dex, France} 
\author{M.J.Ison} %
\affiliation{LPC Caen (IN2P3 - CNRS / EnsiCaen et Universit\'{e}), F-14050
Caen C\'{e}dex, France} 
\affiliation{ Departamento de F\'{i}sica, Facultad de Ciencias Exactas y Naturales, Universidad de Buenos Aires, Pabell\'on $I$, Ciudad Universitaria, $1428$, 
Buenos Aires, Argentina.}
\author{C.O.Dorso} %
\affiliation{ Departamento de F\'{i}sica, Facultad de Ciencias Exactas y Naturales, Universidad de Buenos Aires, Pabell\'on $I$, Ciudad Universitaria, $1428$,
 Buenos Aires, Argentina.}
\date{\today}

\begin{abstract} An information theory description of finite systems 
explicitly evolving in time is presented. 
We impose a MaxEnt variational principle 
on the Shannon entropy at a given time while the constraints are 
set at a former time. The resulting density matrix 
contains explicit time odd components in the form of collective flows.
As a specific application we consider the dynamics of the expansion
in connection with heavy ion experiments.
Lattice gas and classical molecular dynamics simulations
are shown.
\end{abstract}

\pacs{25.70 -z, 25.70.Mn, 25.70.Pq, 02.70.Ns
 	}

\keywords{ Information theory; Collective flow; Multifragmentation
 }
 
\maketitle

\section{Introduction}\label{intro}

The microscopic foundations of thermodynamics are well 
established using the Gibbs hypothesis of statistical 
ensembles maximizing the Shannon entropy\cite{jaynes}. 
At the thermodynamic limit, the various Gibbs
ensembles converge to an unique 
thermodynamic equilibrium. However, most of the systems 
studied in physics do not correspond to this limit\cite{hill}.
In finite systems the various Gibbs ensembles 
are not equivalent\cite{inequiv} and the physical meaning 
and relevance of these different equilibria has to be
investigated.  

A common interpretation of a statistical ensemble for 
a finite system is given 
by the Boltzmann ergodic assumption. 
In this interpretation the statistical ensemble represents 
the collection of successive snapshots of a physical system 
evolving in time, and the state variables are identified 
with the conserved observables.  
This interpretation suffers from important drawbacks. First,  
even for a truly ergodic Hamiltonian, a finite time experiment 
may very well achieve ergodicity only on a subspace of the total 
accessible phase space\cite{thirring}. Moreover, ergodicity applies 
to confined systems and thus it requires the definition of boundary 
conditions 
. Then the statistical ensemble directly depends on the boundary 
conditions and we will discuss 
that an exact knowledge of the boundary corresponds to an infinite 
information and is therefore hardly compatible with the very 
principles of statistical mechanics
. Finally, the systems experimentally accessible are often not 
confined but freely evolve in the vacuum, as it is notably the 
case for 
heavy ion collisions. 
The concept of a stationary equilibrium
defined by the variables conserved by the dynamics in a
hypothetical constraining box, is not useful for these 
systems.

However statistical approaches, expressing the reduction 
of the available information to a limited number of collective 
observables, are still pertinent to complex 
systems even if the dynamics does not allows at any 
time a total and even exploration of the energy shell\cite{balian}. 
In this interpretation the MaxEnt postulate 
has to be interpreted as a minimum information postulate 
which finds its justification in the complexity of the dynamics 
independent of any time scale\cite{jaynes,balian}.

This information theory approach is a very powerful extension 
of the classical Gibbs equilibrium: any arbitrary observable 
can act as a state variable, 
and all statistical quantities 
can be unambiguously defined for any number of 
particles\cite{noi}. The price to be paid for such a
generalization is that the constraining variables as well 
as the density matrix continuously evolve in time. 
The time dependence of the process
naturally leads to the appearance of new time-odd constraints 
or collective flows. In the case of an ideal gas of particles 
or clusters we will show that the ensemble of constraints 
forms a closed algebra and that the information at the initial
time is sufficient to calculate the exact density matrix at any 
successive time. 

\section{Statistical equilibria}

When the system is characterized by $L$ observables 
known in average $<\hat{A}_{\ell }>=\mathrm{Tr}\hat{D}
\hat{A}_{\ell }$, statistical equilibrium corresponds to 
the maximization of the constrained entropy

\[
S_c=-\mathrm{Tr}\hat{D}\log \hat{D}
-\sum_{\ell }\lambda _{\ell }<\hat{A}_{\ell }> 
\]

where $\hat{D}=\sum_{\left( n\right) }\left| 
\Psi ^{\left( n\right) }
\right\rangle\;p^{\left( n\right) }\;\left\langle 
\Psi ^{\left( n\right) }\right| $ is the density 
matrix, and $\vec{\lambda }=\{\lambda _{\ell }\}$ 
are Lagrange multipliers. Gibbs equilibrium
is then given by 

\begin{equation}
\hat{D}_{\vec{\lambda }}=\frac{1}{Z_{\vec{\lambda }}}\exp -\vec{
\lambda} .\vec{\hat{A}},  \label{EQ:D}
\end{equation}

where 
$Z_{\vec{\lambda }}$ is the associated partition sum.
It should be noticed that microcanonical thermodynamics\cite{gross} 
can also be obtained from the variation of the Shannon entropy 
in the special case of a fixed energy subspace.
In this case the maximum of the Shannon entropy can be identified
with the Boltzmann entropy 
$max\left( S\right) =\log W\left( E\right)$,
where $W$ is the total state density with the energy E. 
In the following we shall confine ourselves  to 
the Gibbs formulation (\ref{EQ:D}), 
which is more general than the microcanonical 
ansatz. Indeed the microcanonical
density matrix corresponds to an even occupation 
of the whole energy shell while non ergodic components 
can be already included within the Gibbs 
formalism through the introduction of extra constraints.  


\subsection{Boundary condition problem in finite systems}
  
The statistical physics formalism recalled above is valid 
for any system size 
. However, as soon as one $\hat{A}_{\ell }$ 
contains differential operators such as a kinetic energy, 
eq. (\ref{EQ:D}) is not defined,
unless boundary conditions are specified. 
Only at the thermodynamic limit boundary conditions 
are irrelevant, as only in this limit surface effects
are negligible. 
The definition of any density with a finite number of particles
requires the definition of a finite volume. To this aim,
a fictitious container is generally introduced\cite{bondorf}. 
The volume and shape of this unphysical box has no 
influence on the thermodynamics of self-bound systems,
but in the presence of continuum states the situation is different.
Let us consider the standard case  
of the annulation of the wavefunction on the surface $S$ 
of a containing box $V.$ Introducing the projector, $\hat{P}_{S}$ , 
over $S$ and its exterior, the boundary conditions 
reads $\hat{P}_{S}\left| \Psi ^{\left( n\right) }\right\rangle
=0$ for all microstates $(n)$.
Using $\hat{P}_{S}^{2}=\hat{P}_{S}$, 
we can see that this condition 
imposes an extra constraint to the statistical ensemble 
$<\hat{P}_{S}>=\mathrm{Tr}\hat{D}\hat{P}_{S}=0$.
The density matrix then reads

\begin{equation}
\hat{D}_{\vec{\lambda }S}=\frac{1}{Z_{\vec{\lambda }S}}\exp -\vec{%
\ \lambda} .\hat{\vec{A}}-b\hat{P}_{S}  \label{modif-D-2}
\end{equation}

which shows that the thermodynamics of the system 
depends on the whole
surface $S$. For the very same global features such as the
same average particle density or energy, we will have 
as many different thermodynamics as boundary conditions.
More important, to specify the density matrix, the projector 
$\hat{P}_{S}$ has to be exactly known and this is in fact impossible. 
The nature of $\hat{P}_{S}$ is intrinsically different from the usual 
global observables $\hat{A}_{\ell }$. Not only it is a many-body 
operator, but $\hat{P}_{S}$ requires the
exact knowledge of each point of the boundary surface while no or few
parameters are sufficient to define the $\hat{A}_{\ell }.$
This infinity of points corresponds to an infinite amount of 
information to be known to define the density matrix (\ref{modif-D-2}).
This requirement is in contradiction with the statistical mechanics 
principle of minimum information. Thus eq.(\ref{modif-D-2}) 
is unphysical, and the same is true for the standard $(N,T,V)$
or $(N,E,V)$ ensembles when dealing with finite unbound unconfined 
systems.

\subsection{Incomplete knowledge on the boundaries}

One way to get around the difficulties encountered to take into
account our incomplete knowledge on the boundaries,  
is to introduce a hierarchy of observables describing the size and
shape of the matter distribution.

For example, if only the average system size $<\hat{R}^{2}>$ 
is known, 
the minimum information principle implies

\begin{equation}
\hat{D}_{\beta ,P}=\frac{1}{Z_{_{\beta ,P}}}
\exp -\beta \left( \hat{H}+PR_{0}%
\hat{R}^{2}\right) ,  \label{EQ:D-T-P}
\end{equation}

which is akin an isobar canonical ensemble, since the 
additional Lagrange multiplier $\lambda _{R^{2}}$ 
imposing the size information
has the dimension of a pressure when divided by 
a typical scale $R_{0}$ and by the temperature, 
$\lambda _{R^{2}}=\beta PR_{0}$.

A typical application of this concept is the so called 
freeze-out hypothesis in nuclear collisions 
: at a given time $t_0$ the main
evolution (i.e. the main entropy creation) is assumed to stop
and partitions are supposed
to be essentially frozen. 
Typically thermal and chemical equilibrium is assumed, 
meaning that the information at $t_0$ on the energetics 
and particle numbers is limited to the observables $<\hat{H}>$
and $<\hat{N}_{f}>$ for the different species $f$ 
\cite{high:energy,bondorf}.
Freeze-out occurs when the system has expanded to a finite size.
Then at least one measure of the system's compactness
should be included 
. The limited knowledge of the system extension
leads to a minimum biased density matrix given by eq.(\ref{EQ:D-T-P})
\cite{footnote1}.
 

\section{Multiple time statistical ensembles}

As soon as one of the constraining observables 
$\hat{A}_{\ell }$ is not a constant of the motion, 
the statistical ensemble (\ref{EQ:D}) is not stationary. 
A single time description may still looks appropriate in the freeze out 
configuration discussed in the last section. Indeed
in many physical cases one can clearly identify a specific time
at which the information concentrated in a given
observable is frozen (i.e. the observable expectation value 
ceases to evolve). 
However this freeze out time is in general fluctuating
and different for different observables. For example for the
ultra-relativistic heavy ion reactions two freeze-out times 
are discussed\cite{high:energy}, one for the chemistry 
and one for the thermal agitation.
We need therefore to define a statistical ensemble
constrained by informations coming from different times.

Let us now suppose that the different informations on the system, 
$<\hat{A}_{\ell }>$, are known at different times $t_{\ell }$: 
$<\hat{A}_{\ell }>_{t_{\ell }}=\mathrm{Tr}\hat{D}
\left( t_{\ell }\right) \hat{A}_{\ell }$.
A generalization of the Gibbs idea is that at a time $t$ 
the least biased state of the system is the maximum 
of the Shannon entropy, considering all informations as constraints.  

The maximization of the entropy at time $t$ with the various 
constraints $<\hat{A}_{\ell }>_{t_{\ell }}$ known at former times 
$t_{\ell }$ corresponds to the free maximization of 

\begin{equation}
S_c =-\mathrm{Tr}\left( \hat{D}\left( t\right) 
\log \hat{D}\left(t\right) +\sum_{\ell =1}^{L}\lambda _{\ell }
\hat{A}_{\ell }\hat{D}\left(t_{\ell }\right) \right) ,  
\label{EQ: S-Two-times}
\end{equation}

where the $\lambda _{\ell }$ are the Lagrange parameters 
associated with all the constraints. This maximization will 
lead to a density matrix which can be considered as a generalization 
to time dependent processes of the Gibbs
ensembles (\ref{EQ:D}).

Let us consider the case of a deterministic evolution
\begin{equation}
\partial _{t}\hat{D}=\{\hat{H},\hat{D}\},  \label{EQ:Liouville}
\end{equation}
where 
$\{.,.\}$ are Poisson bracket
in classical physics and commutators divided by $i\hbar $ in quantum
physics. 
The minimum biased density matrix is
given by\cite{annals} 

\begin{equation}
\hat{D}_{\vec{\lambda }}\left( t\right) =\frac{1}{Z_{\vec{\lambda }%
}\left( t\right) }\exp -\vec{\ \lambda .\hat{A}}^{\prime }[\hat{D}_{%
\vec{\lambda }}\left( t\right) ],  \label{EQ:D-equil-t-tprime}
\end{equation}

where the $\hat{A}_{\ell }^{\prime }$ represent the time evolution
of the constraining observables $\hat{A}_{\ell }$ in the Heisenberg
representation $\hat{A}_{\ell }^{\prime }\mathcal{[}\hat{D}
\left( t\right) ]=\hat{A}_{\ell }^{\prime }\left( \Delta t_{\ell }
\right) =e^{-i\Delta t_{\ell}\hat{H}}\hat{A}_{\ell }
e^{i\Delta t_{\ell }\hat{H}}$:

\[
\hat{A}_{\ell }^{\prime }
=\hat{A}_{\ell
}+\sum_{p=1}^{\infty }\frac{(t-t_{\ell })^{p}}{p!}
\hat{B}_{\ell }^{(p)}
\]
\[
\;\;  \;\; 
\hat{B}^{(p)}=\{\hat{H},\hat{B}^{(p-1)}\} \;\; ; \;\;
\hat{B}^{(0)}=\hat{A}.
\]


Eq. (\ref{EQ:D-equil-t-tprime})
can be interpreted as the introduction of additional 
constraints $\hat{B}_{\ell }^{(p)}$ and additional 
Lagrange parameters $\nu _{\ell }^{(p)}$
associated with the time evolution of the system 

\begin{equation}
\hat{D}_{\vec{\lambda ,\nu }}=\frac{1}{Z_{\vec{\lambda ,\nu }}}\exp -%
\vec{\ \lambda .\hat{A}-}\sum_{p=1}^{\infty }\vec{\nu }^{(p)}\vec{.%
\hat{B}}^{(p)}[\hat{D}_{\vec{\lambda ,\nu }}],  \label{multistep}
\end{equation}

Eq.(\ref{multistep}) is an exact solution of
the complete many body evolution problem eq.(\ref{EQ:Liouville}) 
with a minimum information hypothesis on the final time $t$ 
having made few observations $<\hat{A}_{\ell }>$ at previous times 
$t_{\ell }$, which shows the wide domain
of applicability of information theory. 
A generalization of this theory to non deterministic evolutions
can be found in ref.\cite{annals}.
We can see from eq.(\ref{multistep})
that in general an infinite amount of information, 
i.e. an infinite number of Lagrange multipliers are 
needed if we want to follow the system evolution 
for a long time. 
However, different interesting physical
situations exist, for which the series can be analytically summed up. 
In this case, a limited information 
(the knowledge of a small number of average
observables) will be sufficient to describe the whole 
density matrix at any time, under the unique hypothesis 
that the information was finite at a given time.

\section{The dynamics of the expansion}\label{dynamics}

Let us now apply the above formalism to 
transient unconfined  systems.
Let us concentrate on a scenario often
encountered experimentally: 
a finite system of loosely interacting particles with a 
finite extension in an open space. We shall  
assume that at a given freeze out time $t_{0}$
the system can be modelized as a non interacting ensemble 
of $n=1,\dots,N$ particles or fragments, and  
a definite value for the mean square radius $<\hat{\vec{R}}^{2}>$ 
(with $\hat{\vec{R}}^{2}=\sum_{n}\hat{\vec{r}}_{n}^{2}$) 
characterizes the ensemble of states.
Then we have to introduce the constraining observable 
$\hat{A}_{2}=\vec{\hat{R}}^{2}$ associated with a 
Lagrange multiplier $\lambda _{0}$.
If time is not taken into account, the maximum entropy solution 
is given by 

\begin{equation}
\hat{D}_{\beta \lambda _{0}}=\frac{1}{Z_{_{\beta \lambda _{0}}}}\exp -\beta
\sum_{n}\left( \frac{\hat{\vec{p}}_{n}^{2}}{2m}+\frac{\lambda _{0}}{\beta 
}\hat{\vec{r}}_{n}^{2}\right) .  \label{EQ:D-osc}
\end{equation}

Eq.(\ref{EQ:D-osc}) is akin to a system of non-interacting particles 
trapped in a harmonic oscillator potential with a string 
constant $k=2\lambda _{0}/\beta .$ From
the partition sum, the EOS are easily derived: 

\[
<\hat{\vec{p}}_{n}^{2}>=\frac{3m}{\beta }\;\;\;;\;\;\;<\hat{\vec{r}}
_{n}^{2}>=\frac{3}{2\lambda _{0}}. 
\]

Since $\lambda _{0}\hat{\vec{R}}^{2}$ is not an external 
confining potential but only a finite size
constraint, the minimum biased distribution (\ref{EQ:D-osc}) 
is not stationary.  
To take into account the time
evolution, we must introduce additional constraining 
observables 

\[
\hat{B}_{R}^{(1)} 
=-\sum_{n}\frac{1}{m}
\left( \hat{\vec{p}}_{n}\cdot \hat{\vec{r}}_{n}+\hat{\vec{r}}%
_{n}\cdot \hat{\vec{p}}_{n}\right) \;\; ; \;\;
\hat{B}_{R}^{(2)} 
=\sum_{n}\frac{2\vec{\hat{p}}_{n}^{2}}{m^{2}}.
\]

Since $\{\hat{H},\hat{B}_{R}^{(2)}\}=0$, all the other 
$\hat{B}_{R}^{(p)}$
with $p>2$ are zero. 
The density matrix is given by
\begin{eqnarray}
& \hat{D}_{\beta ,\lambda _{0}}(t)=\frac{1}{Z_{\beta ,\lambda _{0}}}\exp
\sum_{n}-\beta_{eff}\left( t\right) \frac{\hat{\vec{p}}_{n}^{2}}{2m}
-\lambda _{0}\hat{\vec{r}}_{n}^{2} \nonumber \\
& +\frac{\nu _{0}\left( t\right) }{2}
\left( \hat{\vec{p}}_{n}\cdot \hat{\vec{r}}_{n}+\hat{\vec{r}}%
_{n}\cdot \hat{\vec{p}}_{n}\right), \nonumber \\ \label{EQ:GP-expan}
\end{eqnarray}
with 
\begin{equation}
\beta_{eff}\left( t\right) =\beta +2\lambda _{0}\left( t-t_{0}\right)
^{2}/m  \;\; ;\;\; \nu _{0}\left( t\right) 
=2\lambda _{0}\left( t-t_{0}\right) /m \label{EQ:beta-t-}
\end{equation}


The density matrix (\ref{EQ:GP-expan}) can be
interpreted as a radially expanding ideal gas. 
Indeed the distribution can be written as 

\begin{eqnarray}
\hat{D}_{\beta ,\lambda _{0}}(t) & = & \frac{1}{Z_{\beta ,\lambda _{0}}}\exp
\sum_{n}-\beta_{eff}\left( t\right) \frac{\left( \hat{\vec{p}}
_{n}-m\alpha\left( t\right) \hat{\vec{r}}_{n}\right) ^{2}}{2m} \nonumber \\ 
& & -\lambda_{eff}\left( t\right) \hat{\vec{r}}_{n}^{2} \nonumber \\  
\label{EQ:D-expan}
\end{eqnarray}

where $\alpha=\nu _{0}\left( t\right) /
\beta_{eff}\left( t\right)$ represents a Hubblian factor  
and the confining Lagrange multiplier is transformed into 

\begin{equation}
\lambda_{eff}\left( t\right) =\lambda_{0}-\frac{m}{2}\frac{\nu
_{0}^{2}\left( t\right) }{\beta_{eff}\left( t\right) }=\frac{\lambda
_{0}\beta m}{\beta m+2\lambda_{0}\left( t-t_{0}\right) ^{2}}
\label{EQ:lambda-prime}
\end{equation}

The term $m\alpha\left( t\right)\hat{\vec{r}}_{n}$ correcting 
the momentum in eq.(\ref{EQ:D-expan}) can be interpreted as a
momentum produced by a radial velocity 
$\alpha\left( t\right) \hat{\vec{r}}_{n}$. 
This proportionality of the velocity with $\hat{r}_{n}$ 
shows that the motion is self similar. As a consequence, 
when this collective motion is subtracted from the 
particle momentum, the density matrix (\ref{EQ:D-expan}) 
corresponds at any time to a standard equilibrium 
(\ref{EQ:D-osc}) in the local rest frame. 
 
In this case the infinite information which is a priori needed 
to follow the time evolution of the
density matrix according to eq.(\ref{multistep}), reduces to the three
observables $\hat{\vec{r}^{2}}$, $\hat{\vec{p}^{2}}$, $\hat{\vec{r}}
\cdot \hat{\vec{p}}+\hat{\vec{p}}\cdot \hat{\vec{r}}$. 
Indeed these operators form a closed Lie algebra, and
the exact evolution of 
(\ref{EQ:D-expan}) preserves it algebraic structure.


The description of the time evolution when describing unconfined 
finite systems has introduced a new phenomenon: the expansion. 
One should then consider a more general equilibrium of a finite-size 
expanding finite-systems with $\beta ^{\prime }$, $\alpha$ and 
$\lambda _{0}^{\prime }$ as free parameters. Then, 
if the observed minim biased distribution at time $t$ is coming 
from a confined system at time $t_{0}$, the three parameters 
$\beta ^{\prime }$, $\alpha$ and $\lambda _{0}^{\prime }$ should 
be linked to the time $t_{0}$ the initial temperature $\beta ^{-1}$ 
and the initial $\lambda _{0}$ by equations (\ref{EQ:beta-t-}) 
and (\ref{EQ:lambda-prime})\cite{npa}.

The important consequence is that radial flow is a necessary
ingredient of any statistical description of unconfined finite systems: the
static (canonical or microcanonical) Gibbs ansatz in a confining 
box which is often employed\cite{bondorf} misses this crucial point. 
On the other hand, if a radial flow is observed in the experimental 
data, the formalism we have developed allows to associate this 
flow observation to a distribution at a former time when flow was 
absent. This initial distribution corresponds to a standard static 
Gibbs equilibrium in a confining harmonic potential, i.e. 
to an isobar ensemble.

\section{Numerical simulations}

\begin{figure}[htbp]
   \setlength{\abovecaptionskip}{-10pt}
   \centering
   \includegraphics[width=8cm,angle=0,clip=]{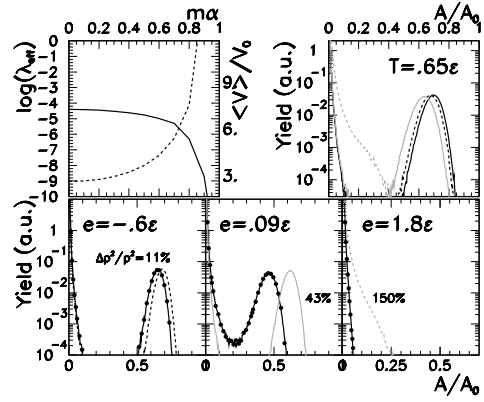}%
   \caption{Upper left: effective pressure (full line) and average
volume normalized to the ground state volume 
$V_{0}=A$ (dashed line) as a function of the collective radial flow. 
Upper
right and lower part: fragment size distributions in the expanding Lattice
Gas model. Distributions without flow (full black lines) are compared with
distributions with 11\% (dashed black), 43\% (full grey) and 150\% (dashed
grey) contribution of radial flow at the same temperature (upper right) and
at the same total energy (lower part).}
\label{fig1}
\end{figure}
 
As we have already mentioned in section \ref{dynamics},
in the hypothesis of negligible interaction between the system's
constituents the expansion is self-similar, implying that  
the situation is equivalent to a standard Gibbs equilibrium 
in the local rest frame. 
In the expanding ensemble the total average kinetic energy 
per particle is simply the sum of the thermal energy
$ \langle e_{th}\rangle= {3}/({2\beta })$ and the radial flow 
$\langle e_{fl}\rangle =m\alpha^{2}\langle r^{2}\rangle/2$.  

This scenario is often invoked 
in the literature \cite{bondorf} to justify the treatment 
of flow as a collective radial velocity superimposed on thermal 
motion; however eq.(\ref{EQ:D-expan})contains also an
additional term $\propto r^{2}$ which corresponds to an outgoing
pressure 
.
The phase diagram and fragment observables are therefore 
expected to be modified by the presence of flow even in the 
self similar approximation.
To quantify this statement, we have performed calculations
in the Lattice Gas Model\cite{npa}, and the results are shown
in Figure \ref{fig1}.  

The effective pressure $\lambda _{eff}$ as well as the 
associated average volume are shown in the
upper left part of figure \ref{fig1} as a function 
of the collective radial flow for a given pressure 
and temperature 
. The Lagrange parameter $\lambda _{eff}$ being a
decreasing function of $\alpha $, 
the critical point is moved towards higher pressures 
in the presence of flow \cite{jou}. 
However one can see that the effect is very
small up to $m\alpha \approx .6$ (which corresponds to 
$\Delta p^2/p^2 = \langle e_{fl}\rangle /\langle e_{th}\rangle 
\approx 67\%$
flow contribution). 
In this regime the cluster size distributions displayed 
in the upper right part of figure \ref{fig1} are
only slightly affected. On the other side if collective 
flow overcomes a threshold value $\Delta p^2/p^2\approx 100\%$ 
the average volume 
shows an exponential increase and the outgoing flow pressure
leads to a complete fragmentation of the system (dashed grey line in 
the lower part of fig.\ref{fig1}).
We can also observe that
an oriented motion is systematically less effective than 
a random one to break
up the system. This is shown in the lower part of figure \ref{fig1} 
which compares for a given $\lambda$ distributions 
with and without radial flow at the same
total deposited energy: for any value of radial flow equilibrium 
in the standard microcanonical ensemble corresponds to more 
fragmented configurations.


\begin{figure}[htbp]
   \centering
   \includegraphics[width=7cm,angle=0,clip=]{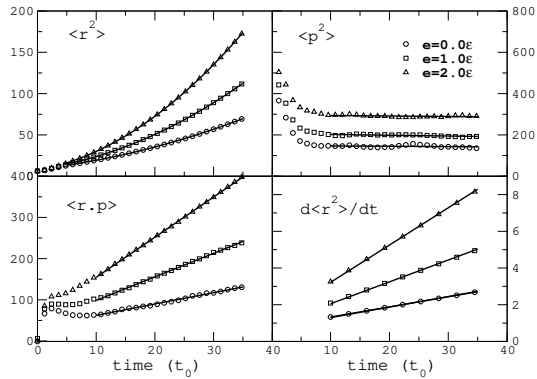}%
   \caption{ Time evolution of $<R^2>$, $<P^2>$ and $<R\cdot P>$
for an initially constrained Lennard Jones system of 147 particles
freely expanding in the vacuum, at different total energies.
Lower right: expansion dynamics (symbols) compared to the prediction
of eq.(\protect\ref{EQ:GP-expan}) (lines).
 }
\label{fig2}
\end{figure}

The above results are relevant for the experimental situation
if and only if the inter-fragment interactions can be 
neglected when the system is still compact enough
to bear pertinent information on the phase diagram.
Indeed only in this case the series (\ref{multistep})
can be analytically summed up and the expansion dynamics 
can be reduced to a self similar flow\cite{annals}.
The validity of the ideal gas approximation eq.(\ref{EQ:GP-expan})
for the expansion dynamics is tested in Figure \ref{fig2}\cite{matias} 
in the framework of classical molecular dynamics\cite{dorso}.
A Lennard Jones system is initially confined in a small volume
and successively freely expanding in the vacuum.
We can see that after a first phase of the order of $\approx 10$
Lennard Jones time units, where interparticle interactions cannot be 
neglected, the time evolution predicted by eq.(\ref{EQ:GP-expan})
is remarkably fulfilled for all total energies.
This result is due to the fact that the system's size and dynamics are
dominated by the free particles, while deviations from a self similar
flow can be seen if the analysis is restricted to bound
particles\cite{matias}. We expect eq.(\ref{EQ:GP-expan}) to 
describe the system evolution even better if the degrees of freedom
$n$ are changed from particles to clusters, as suggested by the Fisher 
model of condensation\cite{fisher}. 
 
\section{Conclusions}\label{concl}
In this paper we have introduced an extension of Gibbs ensembles 
to time dependent constraints
. This formalism gives a statistical description of a system observed  
at a time at which the entropy has not reached its saturating value 
yet, as it may be the case in intermediate energy heavy ion 
reactions\cite{low:energy}. Another physical application 
concerns systems for which the relevant observables pertain 
to different times, as in high energy nuclear
collisions 
\cite{high:energy}.

Our most important result is that any statistical description 
of a finite unbound system must necessarily contain a local 
collective velocity term. Indeed the
knowledge of the average spatial extension of the system 
at a given time, naturally produces a flow constraint 
at any successive time. 
Conversely a collective flow measurement at a given time can be
translated into an information on the system density 
at a former time.
   

\vfill\eject

\begin{thebibliography}{99}  

\bibitem{jaynes}  E. T. Jaynes, \textsl{'Information theory and statistical
mechanics'}, Statistical Physics, Brandeis Lectures, \textbf{vol.3}, 160
(1963).

\bibitem{hill}  T. L. Hill, \textsl{'Thermodynamics of small systems'}, W.
A. Benjamin Inc., New York (1963).

\bibitem{inequiv}  F. Gulminelli, Ph. Chomaz, {\it Phys. Rev.} 
\textbf{E 66} (2002) 46108

\bibitem{thirring}  W. Thirring, H. Narnhofer, H. A. Posch, 
{\it Phys. Rev. Lett.} 
\textbf{91} (2003) 130601.

\bibitem{balian}  R. Balian, 
\textsl{'From microphysics to macrophysics'},
Springer Verlag (1982).

\bibitem{noi}  Ph. Chomaz, F. Gulminelli, \textsl{'Phase transitions 
in finite systems, in Dynamics and thermodynamics of systems 
with long range interactions'}, Lecture Notes in Physics 
\textbf{vol.602}, Springer (2002); F.Gulminelli, Ann.Phys.Fr., in press.

\bibitem{gross}  D. H. E. Gross, 
\textsl{'Microcanonical thermodynamics:
phase transitions in finite systems'}, Lecture Notes 
in Physics \textbf{\
vol.66}, Springer (2001).


\bibitem{bondorf}  J. P. Bondorf et al., {\it Phys. Rep.} 
\textbf{257} (1995) 133; F.Becattini et al., {\it Phys.Rev.} 
{\bf C69} (2004) 024905.

\bibitem{high:energy}  P. Braun-Munzinger et al.,
Invited review for Quark Gluon Plasma 3, 
eds. R. C. Hwa and Xin-Nian Wang, World Scientific Publishing.
 

\bibitem{footnote1}
The ensemble of events dynamically prepared may be 
characterized by an
information about size and shape more complex than 
the simple mean square
root radius. In this case other constraints can 
be introduced such as $\hat{R}^{4}$ if the radii fluctuations 
contain non-trivial information  or 
$\hat{Q }_{2}=2\hat{Z}^{2}-\left( \hat{X}^{2}+\hat{Y}
^{2}\right) $ if the system is not spherical in 
average but has a finite
quadrupole deformation.

\bibitem{annals} Ph.Chomaz, F.Gulminelli and O.Juillet, 
{\it Ann.Phys.},in press.

\bibitem{npa}  F. Gulminelli, Ph. Chomaz, {\it Nucl. Phys.} 
\textbf{734A} (2004) 581.

\bibitem{low:energy}  J. Richert and P. Wagner, {\it Phys. Rep.}
 \textbf{350}(2001) 1. 
 
\bibitem{matias} M.Ison et al., in preparation.

\bibitem{dorso} A. Chernomoretz et al.,
{\it Nucl.Phys.} {\bf A723} (2003) 229. 

\bibitem{fisher}  M. E. Fisher, {\it Physics} 
\textbf{vol. 3}, No.5 (1967) 255.

\bibitem{jou}  D.Jou et al., 
{\it 'Thermodynamics of fluids under flow'}, Springer
(2001) 




 

\end{thebibliography}
\end{document}